\documentclass[sigconf]{acmart}
\usepackage[utf8]{inputenc}
\usepackage{tabularx}
\usepackage{graphicx}
\usepackage{subcaption}
\usepackage{afterpage}
\usepackage{enumitem}
\usepackage{amsmath}
\setlist[itemize]{align=parleft,left=0pt..1em}

\acmConference[WWW ’24 Companion]{ACM Web Conference 2024}{May 13--17, 2024}{Singapore}

\title{AutoML for Large Capacity Modeling of Meta's Ranking Systems}

\author{Hang Yin$^*$,\hspace{1em} Kuang-Hung Liu$^{*}$,\hspace{1em} Mengying Sun,\hspace{1em} Yuxin Chen,\hspace{1em} Buyun Zhang}
\thanks{*These authors contributed equally. $\dagger$ Project lead.}
\author{Jiang Liu,\hspace{1em} Vivek Sehgal,\hspace{1em} Rudresh Rajnikant Panchal,\hspace{1em} Eugen Hotaj,\hspace{1em} Xi Liu}
\author{Daifeng Guo,\hspace{1em} Jamey Zhang,\hspace{1em} Zhou Wang,\hspace{1em} Shali Jiang,\hspace{1em} Huayu Li,\hspace{1em} Zhengxing Chen}
\author{Wen-Yen Chen,\hspace{1em} Jiyan Yang,\hspace{1em} Wei Wen$^{\dagger}$}
\email{{hyin5,khliu,mengyingsun,yuxinc,buyunz,jiangliu,viveksehgal,rudreshp,ehotaj,xliu1}@meta.com}
\email{{dfguo,jameyz,zhouwang,shalijiang,huayuli,czxttkl,wychen,chocjy,wewen}@meta.com}
\affiliation{%
  \institution{Meta Platforms Inc., USA}
  \country{}
}
\date{October 2023}

\renewcommand{\shortauthors}{Yin and Liu, et al.}

\begin{document}

\begin{abstract}
Web-scale ranking systems at Meta serving billions of users is complex. Improving ranking models is essential but engineering heavy. Automated Machine Learning (AutoML) can potentially release engineers from labor intensive work of tuning ranking models; however, it is unknown if AutoML is efficient enough to meet tight production timeline in real-world applications and, at the same time, bring additional improvements to the already strong baselines. Moreover, to achieve higher ranking performance, there is an ever-increasing demand to scale up ranking models to even larger capacity, which imposes more challenges on the AutoML efficiency. The large scale of models and tight production schedule requires AutoML to outperform human baselines by only using a small number of model evaluation trials ($\sim 100$).
This paper presents a sampling-based AutoML search method, focusing on neural architecture search and hyperparameter optimization, with a particular emphasis on addressing aforementioned challenges in Meta-scale production when building large capacity models. Our approach efficiently handles large-scale data demands. It leverages a lightweight predictor-based searcher and reinforcement learning to explore vast search spaces, significantly reducing the number of model evaluations.
Through experiments in large capacity modeling for CTR and CVR applications, we have demonstrated that our method achieves outstanding Return on Investment (ROI) versus human tuned baselines, with up to $0.09\%$ Normalized Entropy (NE) loss reduction or $25\%$ Query per Second (QPS) increase by only sampling one hundred models on average from a curated search space. The proposed AutoML method has already made real-world impact where a discovered Instagram CTR model with up to $-0.36\%$ NE gain (over existing production baseline) was selected for large-scale online A/B test and show statistically significant gain. These production results proved AutoML efficacy and accelerated its adoption in ranking systems at Meta. 
\end{abstract}

\maketitle

\section{INTRODUCTION}

The field of Automated Machine Learning (AutoML) has emerged as a pivotal research area at Meta, driven by the ever increasing complexity and scale of web-based systems, which now serve billions of users. With advancements in hardware, the demand for evolving these vast models has grown exponentially, posing formidable challenges due to constraints in human resources and computational capacities. The goal of AutoML is to automate the process of applying machine learning (e.g., model architecture selection, hyperparameter optimization, etc.) to discover promising models with better Return on Investment (ROI): complex models with accuracy improvement or computationally efficient architecture.
This endeavor has garnered considerable attention from both academia and industry dedicated to the advancement of machine learning and artificial intelligence, like Meta. In this work, we present a sampling-based AutoML search method that focuses on neural architecture search and hyperparameter joint optimization and discuss how our proposed method tackles real-world production challenges:
\begin{itemize}[leftmargin=*]
    \item \textbf{Industry-scale ranking model.} Meta's ranking models continue to grow both in size and complexity with advancements in hardware and the need for serving high quality Click Through Rate (CTR) and Conversion Rate (CVR) prediction. This poses significant challenge for AutoML as increases in model size leads to increases in search space, and increase in model complexity leads to costly model evaluation.
    \item \textbf{Tight production schedule.} In real-world application, production ranking model iteration follows tight development and deployment schedule to ensure the release of highest-quality model. The strict production schedule leaves AutoML limited window of time to search for strong candidates.
    \item \textbf{Compatibility with human baselines and evolving stack.} Meta's ranking models are developed by many in-house engineers and undergoes continuous evolution. While it's widely recognized that scaling-up large capacity models is a labor intensive effort and a common pain point for engineers, the strong and dynamic baselines established by expert human sets a high bar for AutoML. Compatibility of search space to human ideas and compatibility of discovered model to serving stack are essential for AutoML to deliver meaningful end to end impact.
    \item \textbf{Limited computation resource.} The challenge of AutoML lies in the time it takes to discover and validate a new model; during this process, it relies on heavy computation to enable automation. However, the increase in computing power required to run AutoML can incur additional (sometime significant) cost. AutoML needs to demonstrate high ROI (not only effective but also efficient) to justify its deployment in production.
\end{itemize}

\noindent In our pursuit of industry-level AutoML at Meta scale, we improved a predictor-based searcher~\cite{wen2020neural}.
It serves as an efficient sampling method by mapping model configurations to performance proxies, obviating the need for real-data model evaluations. 
This State-of-the-Art (SOTA) method is highly flexible and fully parallelizable. Its compatibility to human baselines and serving stack is also high.
To deal with the significant challenges we mentioned above, we advance this method by:
\begin{itemize}[leftmargin=*]
    \item \textbf{Low-fidelity Evaluation with High Ranking Quality.} Sampling-based searchers often struggle to gather enough samples to ensure the generalization capacity of the searcher. To overcome this, we employ low-fidelity evaluation to expedite data collection by training a small portion of data. We first evaluate a subset of randomly selected models and ensure their training fully converged (typically require large amount of data). By conducting ranking correlation analysis, we determine the potential for reducing data while maintaining similar ranking quality. This enables us to evaluate subsequent models with reduced data, saving valuable computation resources. Moreover, the dataset of random models we trained can be reused to warm start our searcher.
    This trait is given by the searcher's ability to reuse historical data.
    \item \textbf{Training Curve Fitting.} Our use of low-fidelity evaluation favor models with short-term performance and can sometimes miss models with long term gain but only mediocre short-term performance. To address this challenge, we propose a new proxy that considers both short-term loss and training curve trends. The proposed proxy helps us identify models with both strong short-term performance and favorable long-term trends.
    \item \textbf{Predictor Optimization with Pairwise Ranking Loss.} Even with low-fidelity evaluation, evaluating large number of models is expensive and in practice we only have access to a very small number of examples (denoted by $N$) to train a predictor. To alleviate label scarcity, we use pairwise ranking loss~\footnote{\url{https://pytorch.org/docs/stable/generated/torch.nn.MarginRankingLoss.html}} to train the predictor. This increases the number of labels to $\mathcal{O}(N^2)$, and is proven more accurate than mean squared error loss.
    \item \textbf{Ensemble Modeling and Multi-task Learning.} Sampling-based searchers often suffer from poor generalization under limited sampling points. Further compounding this challenge is the noisy and non-stationary production data. To address these challenges, we employ ensemble modeling and multi-task learning for the predictor to improve its generalization capabilities.
    \item \textbf{Predictor and RL Searcher Integration.} To find the best-performing model over a huge search space with an efficient sampler, we adopt the REINFORCE algorithm~\cite{10.1007/BF00992696}, where we learn a stochastic sampling policy to minimize a predefined reward, which combines performance metrics like Normalized Entropy (NE), FLOPs and Queries Per Second (QPS). The efficiency of the RL sampler allows us to greatly reduce the number of models that need to be evaluated compared to a random sampler, while also identifying promising models that satisfy different production requirements. 
\end{itemize}

\noindent With the proposed predictor-based RL method, we find that
\begin{itemize}[leftmargin=*]
  \item It can quickly facilitate large capacity modeling with better ROI than manual tuning. Compared to strong human baselines, we demonstrate the proposed method can bring up to $0.09\%$ Normalized Entropy (NE)~\footnote{Normalized Entropy (NE)~\cite{he2014practical} is the optimization objective used at Meta for ranking systems.} loss reduction or $25\%$ Query per Second (QPS) increase by only sampling one hundred models on average from a curated search space.
  \item It is able to discover promising model architectures at different levels of training/serving infrastructure capacity effectively.
  \item It outperforms other sampling-based algorithms such as random search and Bayesian optimization.
\end{itemize}

\section{RELATED WORK}

Various deep models have been deployed for ads ranking service to generate high quality CTR and CVR predictions \cite{10.1145/2988450.2988454, DBLP:journals/corr/GuoTYLH17, zhou2018deep, Huang_2019}. A main component of ranking model is the feature interaction module which are used to model the relationship among the feature embeddings. Multiple studies have shown a better designed feature interaction module can significantly improve model performance \cite{10.1145/2988450.2988454, DBLP:journals/corr/abs-1708-05027, 10.1145/3447548.3467133, 10.1145/3219819.3220023, DBLP:journals/corr/abs-2003-11235, DBLP:journals/corr/abs-1810-11921}. More recently, motivated by the architecture advancement in computer vision and NLP, SOTA ranking models also adopt a deep stacking structure that is composed of repeating blocks of the same interaction module \cite{zhang2022dhen, naumov2019deep}. As the model grow more complex, number of architecture decisions increases and training stability become more sensitive to training hyperparameters and thus require careful fine tuning. Our goal is to apply AutoML to automatically optimize for these modeling decisions.

AutoML aims to automate various stages of the machine learning process: data preparation, feature engineering, model generation, and model evaluation \cite{HE2021106622, 2018arXiv181013306Y, DBLP:journals/corr/abs-1904-12054, DBLP:journals/corr/abs-1906-02287}. In this work, we focus on the automation of the model generation and model evaluation stages. Hyperparameter optimization (HO) and neural architecture search (NAS) are common optimization methods used in AutoML. Problem wise, HO typically targets tuning training related parameters (e.g., the learning rate, batch size, etc) \cite{Feurer2019, DBLP:journals/corr/ClaesenM15,JMLR:v13:bergstra12a} while NAS focuses on model-related parameters (e.g., the number of layer of NN, size of MLP, etc.) \cite{elsken2019neural, DBLP:journals/corr/abs-1905-01392, DBLP:journals/corr/abs-2006-02903}. In this work, we aim to jointly tune both training-related and model-related parameters. Technique wise, HO and NAS share many underlying search method such as Bayesian optimization (BO) \cite{DBLP:journals/corr/abs-1807-06906, DBLP:journals/corr/abs-1802-07191, swersky2014raiders}, evolution-based algorithm (EA) \cite{pmlr-v70-real17a, elsken2019efficient, DBLP:journals/corr/SuganumaSN17, miikkulainen2017evolving}, reinforcement learning (RL) \cite{DBLP:journals/corr/ZophL16, DBLP:journals/corr/ZophVSL17, pmlr-v80-pham18a}, etc. Many of existing work were evaluated with small curated public datasets whereas our proposed method were designed for challenging real-world production environment.   

\section{METHOD}

At a high level, our proposed AutoML method unfolds in two main parts: train a predictor and sample with RL. The complete workflow is outlined below (also illustrated in Figure~\ref{fig:workflow}):
\begin{figure*}[ht]
  \centering
  \includegraphics[width=0.75\linewidth]{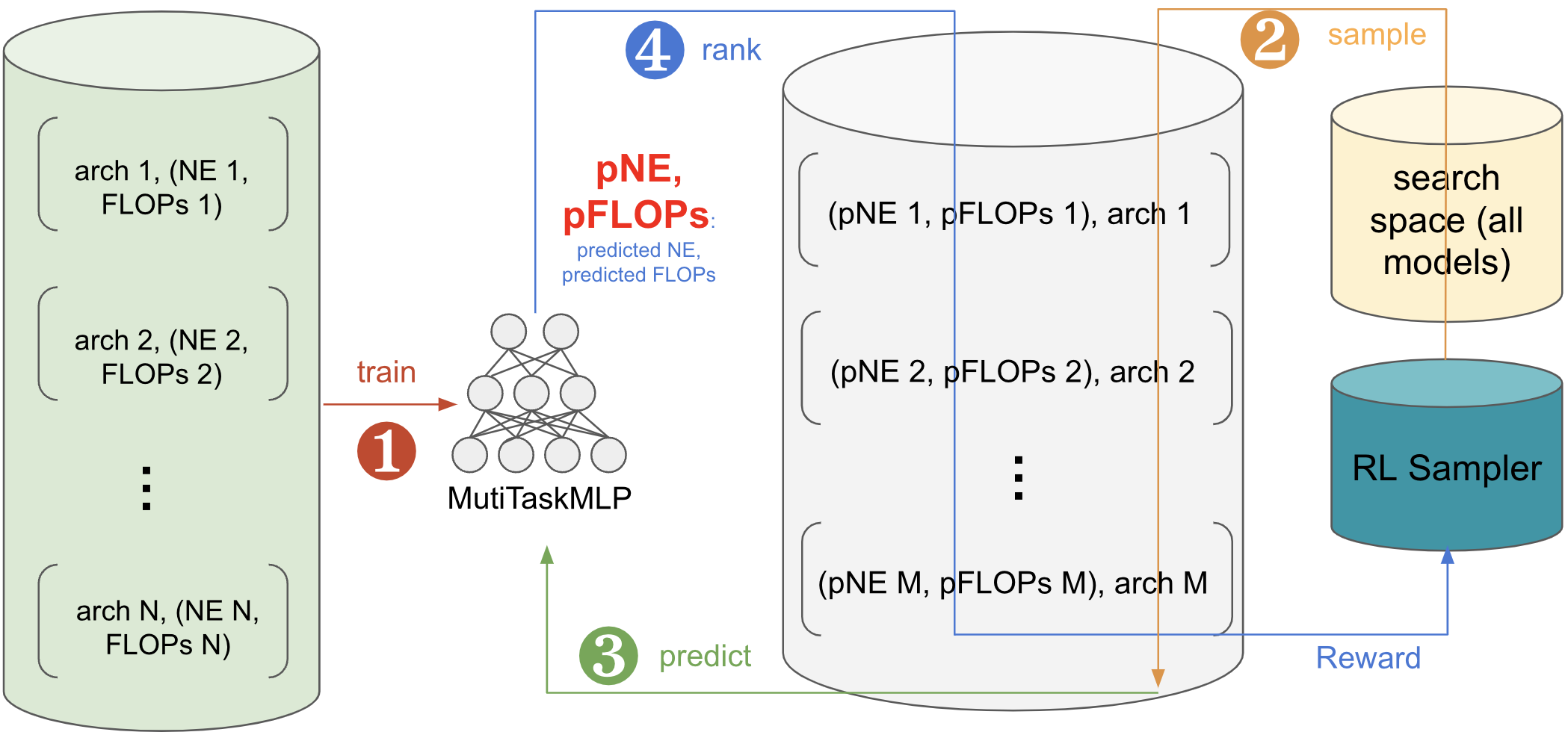}
  \caption{Training and applying the Predictor-based RL searcher. NE and FLOPs are data labels we used to train a predictor. In practice, we apply Eq.~(\ref{eq:long-term}) instead of NE for model accuracy. For sampling architectures, we combine RL with ROI constraint by following Eq.~(\ref{eq:rl_roi}).}
  \label{fig:workflow}
\end{figure*}
\begin{itemize}
    \item Step 1: train a machine learning model (dubbed as ``predictor'') which can predict ``NE gain'' and FLOPs for a given model architecture (``arch'').
    ``NE gain'' is defined as Normalized Entropy~\cite{he2014practical} loss improvement versus a baseline and a negative value implies improvement since a smaller NE loss is better.
    The training dataset include triplets of model arch encoding, NE gain and FLOPs, which are ground-truth labels obtained from historical data or by running model evaluation;
    \item Step 2: a RL agent samples a set of model ``arch'' ;
    \item Step 3: RL sampled models are fed into the predictor to get predicted NE gain (``pNE'') and predicted FLOPs (``pFLOPs'');
    \item Step 4: ``pNE'' and ``pFLOPs'' are sent back to the RL agent as reward, such that it can learn to sample better models.
\end{itemize}
In the following, we delve into each step in more details.  

\subsection{Training the predictor}
\label{sec:train_pred}
The basic idea of the predictor searcher is to build an efficient surrogate model (e.g., MLP), called a "predictor", to map any model configuration to the metrics we are interested in. We then perform model selection based on the predicted metrics and finally validate the top candidates (top-k models) with actual online training. The predictor-based searcher is easily compatible with existing production ranking system stack.
Formally, at Meta, a model \( \pi \) is configured as JSON file, and each model hyperparameter can be converted to one-hot or float encoding. One-hot encoding is used to encode a parameter \( \text{ParamA} \) with \( N \) distinct values as:
\[ \text{Value}_i \rightarrow [0, \ldots, 1, \ldots, 0] \]
where the \(i\)-th position is 1 and all others are 0 if the \(i\)-th value is selected.
For float encoding, each parameter can be mapped to a unique number: $\text{ParamA} \rightarrow \mathbb{R}$.
The final representation of \( \pi \) is the concatenation of all encoding vectors.

After defining the search space and converting model configuration, our next task is to collect samples to train the predictor architecture. 
Real-world production iteration constantly faces strict time constraints and computational resource limitations. Here, we discuss several technical challenges for building an accurate predictor and how we propose to overcome them.

\paragraph{\textbf{Low-fidelity Evaluation with High Ranking Quality.}} Evaluating a large capacity model end-to-end is expensive and therefore it is too costly to generate large amount samples to train a predictor with good generalization performance. As a result, the accuracy of the predicted metrics derived from the predictor (on unseen data) may suffer from limited training samples. To address this, we use low-fidelity evaluation (i.e. early stop training) to save compute cost for each model evaluation such that we can collect more data points under the same computing budget. More specifically, we first evaluate a small set of pilot models (randomly selected), ensuring they are trained on ample data such that their learning curves converge. We then conduct a ranking correlation analysis to determine how much training data can be reduced (i.e., how early training can be stopped) while still preserving these pilot models' relative ranking order based on each model's reduced learning curve. We then apply the same level of training data reduction to all subsequent model evaluation. We call this  process the \emph{low-fidelity evaluation}. Based on our experience, we find that we can consistently achieve 50\% training data reduction from low-fidelity evaluation, i.e., a 50\% compute resource saver for each model evaluation

\begin{figure}
  \centering
  \includegraphics[width=0.8\linewidth]{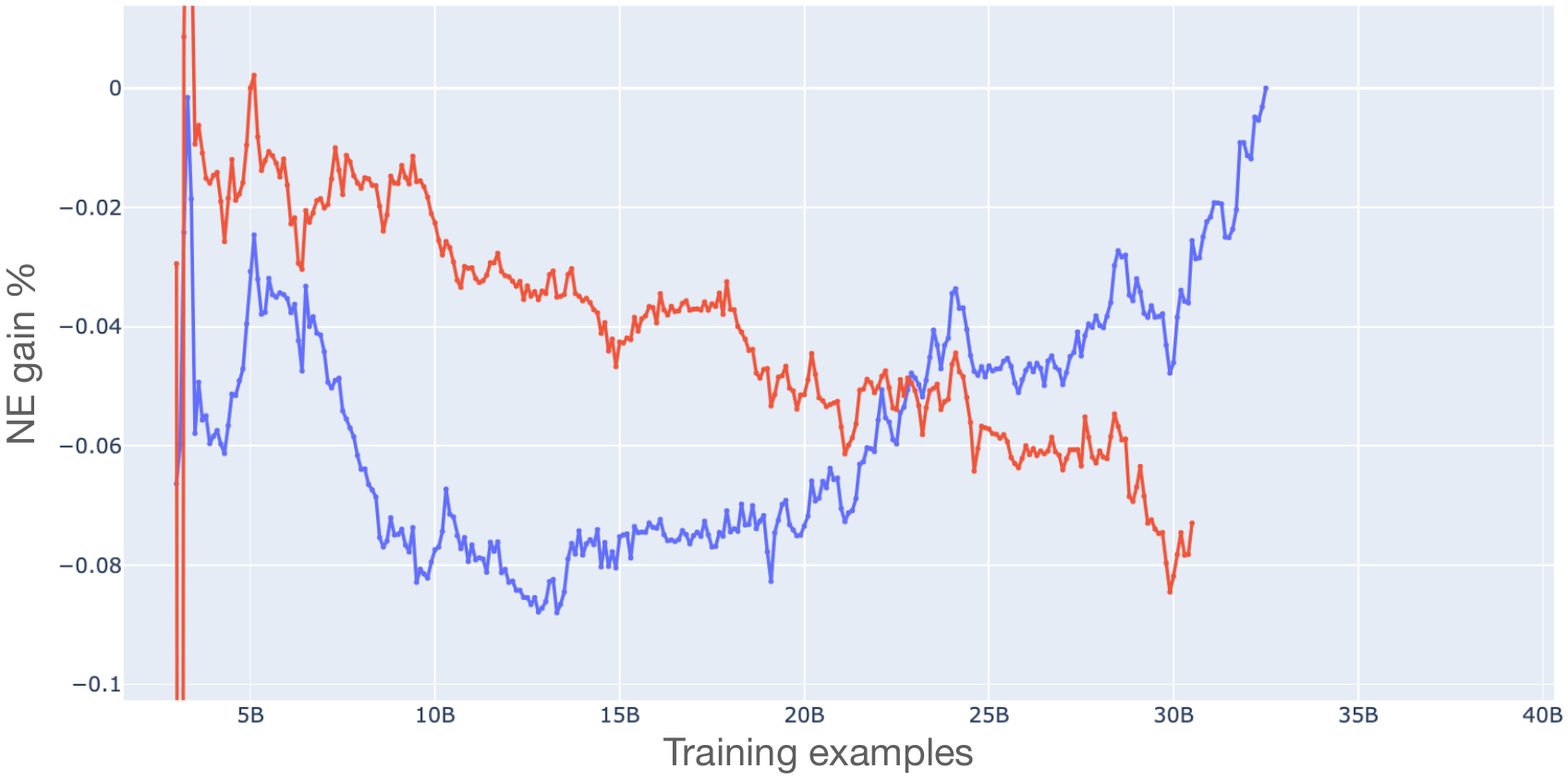}
  \caption{Top models found through different NE metrics (the blue one: raw NE gain evaluated on low-fidelity data; the red one: predicted long-term NE gain calculated by Eq.~\ref{eq:long-term}.}
  \label{fig:long-term}
\end{figure}

\paragraph{\textbf{Training Curve Fitting}.} 
At Meta, building ranking models with long term gain is very critical. While the previously discussed low-fidelity evaluation can bring significant compute saving, selecting model based their reduced learning curve may favor short-term NE gain over long term gain. An example is shown in Fig.~\ref{fig:long-term} where if we naively use raw low-fidelity NE (e.g., short term NE gain at 15B) as the evaluation metric then the blue model is preferred; however, the red model is clearly a better choice long term since it has favorable curve trends (even though it has worse short term NE gain).
To remedy this, we propose a new metric that consider both short-term NE and curve trends. More specifically, we select several points near the end of the low-fidelity evaluation to fit a linear regression model. We then predict the value at a future location based on the linear regression model and use this extrapolated value as the label to train the predictor. 
This approach increases the likelihood of identifying models with both strong short-term performance and favorable long-term trends. Formally, given a short-term NE learning curve, we can represent the predicted long-term NE using the following equations:
\begin{align}
NE(l) &= m \cdot l + c \\
NE_{\text{long-term}} &= NE_{\text{short-term}} + m \cdot \Delta x  \label{eq:long-term}
\end{align}
Where \( NE(l) \) is the fitted NE gain for a given location \( l \) based on the linear regression, \( m \) is the slope of the linear regression, \(NE_{\text{short-term}}\) is the observed NE gain at the curve end, \( \Delta x \) is the step size representing the targeted future location for extrapolation.

\paragraph{\textbf{Training model with Pairwise Ranking Loss}.}
When we use the predictor to estimate the training NE gain of an architecture, to find true top models, our primary focus is its ranking order rather than its proximity to the exact true training NE gain. 
Therefore, we have found that using a pairwise ranking loss is more effective in identifying top models compared to losses such as MSE. Typically, the pairwise ranking loss is represented as follows:
\begin{align}
L(\theta) = \sum_{(i, j) \in D} \max(0, -y_{ij}(f(x_i; \theta) - f(x_j; \theta)) + \epsilon)
  \label{eq:pairwise_rank}
\end{align}
where $L(\theta)$ is the loss function, $(i, j)$ are pairs of samples in the dataset $D$, $y_{ij}$ is the label (indicating which sample in the pair is ``better''), $f(x; \theta)$ is the predictor model, $x_i$ and $x_j$ are the features of samples $i$ and $j$, respectively, and $\epsilon$ is a margin. This formulation assumes that $y_{ij}$ takes a value of $1$ if sample $i$ has higher NE gain than sample $j$ and $-1$ otherwise. The loss is zero if the predicted margin is larger than $\epsilon$, and it increases linearly otherwise.

\paragraph{\textbf{Ensemble Modeling and Multi-task Learning}.} We employ ensemble modeling and multi-task learning to improve predictor's generalization capability. In this work we propose to build \( k \) independently trained predictors where each is a MLP based model that takes a model configuration as input and predicts the corresponding NE gain and FLOPs. Formally, given input encoding $x$, we have a shared neural architecture $NN(x)$ with $L$ layers producing a representation vector $h_L = NN(x)$. For tasks such as prediction of NE gain and FLOPs, the output from the shared architecture is used as input of multiple heads as:
\begin{align}
\begin{split}
NE(x) & = W_{NE} h_L + b_{NE} \\
FLOPs(x) & = W_{FLOPs} h_L + b_{FLOPs}
\end{split}
\end{align}
Finally, the ensemble outputs for the tasks are:
\begin{align}
\begin{split}
NE_{\text{ensemble}}(x) & = \frac{1}{k} \sum_{j=1}^{k} NE^j(x) \\
FLOPs_{\text{ensemble}}(x) & = \frac{1}{k} \sum_{j=1}^{k} FLOPs^j(x)
\end{split}
\label{eq:ensemble}
\end{align}

\subsection{Predictor-based Sampling with Reinforcement Learning}

Previously discussed predictor allows us to easily estimate the performance of any given model without actual training; however, an efficient search method (sampler) is still needed to find an optimum model. In this work, we adopted a reinforcement learning (RL) method---REINFORCE~\cite{10.1007/BF00992696}---to learn a stochastic sampling policy (e.g., learning a configuration sampling distribution) so that the sampled model configuration minimizes some predefined reward. To promote the discovering of models with good performance and cost trade-off, we define the RL reward as a combination of NE gain and cost, e.g., FLOPs. Formally, ROI-constraint search solves the following optimization problem:
\begin{equation}
	\pi^{*} = \text{argmin}_{\pi \in \Pi}(1-\alpha)NE_{\text{ensemble}}(\pi) + \alpha FLOPs_{\text{ensemble}}(\pi),
\label{eq:rl_roi}
\end{equation}
where $NE_{\text{ensemble}}$ and $FLOPs_{\text{ensemble}}$ are generated by a trained predictor, $\pi$ and $\Pi$ represent the model configuration and the corresponding search space, respectively. The coefficient $\alpha$ is a weighting parameter used for NE and cost tradeoffs. 

As well be shown later in our experiment results, we consistently find RL sampler to be efficient (converge within $2000$ steps and only takes several minutes on a single GPU machine) and effective (converged to a model with predicted NE gain that is better than all of the training data) in finding models with strong predicted performance. The highly efficient RL sampler allows us to iterate multiple round of RL search each with a different random seed that can lead to the discovery of a diverse set of local optimal models. It also allows us to explore different part of NE-cost Pareto front by setting a different NE-cost tradeoffs through weighting coefficient $\alpha$. Finally, we select the top-k models (from each round of RL search) for final assessment (long term training) on real data. 



\begin{table*}[t]
\centering
\caption{AutoML search results across different model types}
\label{tab:search_results}
\small
\begin{tabularx}{\textwidth}{>{\centering\arraybackslash}X>{\centering\arraybackslash}X>{\centering\arraybackslash}X>{\centering\arraybackslash}X>{\centering\arraybackslash}X>{\centering\arraybackslash}X}
\hline
\textbf{Model} & \textbf{Method} & \textbf{\# samples} & \textbf{Complexity} & \textbf{3B window NE gain vs baseline} & \textbf{Inference QPS vs manual} \\
\hline
Instagram CTR production & -- & -- & 7x & -- & -- \\
Instagram CTR small & Manual & -- & 18x & -0.07\% & -- \\
Instagram CTR small & AutoML & 25 & 7x & -0.07\% & \textbf{+18.96\%} \\
Instagram CTR small & AutoML & 25 & 13x & \textbf{-0.13\%} & -1.95\% \\
\hline
Instagram CTR scale-up & Manual & -- & 34x & -0.32\% & -- \\
Instagram CTR scale-up & AutoML & 130 & 37x & -0.32\% & \textbf{+24.91\%}\\
Instagram CTR scale-up & AutoML & 130 & 56x & \textbf{-0.36\%} & -8.19\%\\
\hline
Ad CVR production & -- & -- & 7x & -- & -- \\
Ad CVR scale-up & Manual & -- & 38x & -0.24\% & -- \\
Ad CVR scale-up & AutoML & 150 & 30x & -0.24\% & \textbf{+6.50\%}\\
Ad CVR scale-up & AutoML & 150 & 33x & -0.29\% & -5.44\%\\
Ad CVR scale-up & AutoML & 150 & 43x & \textbf{-0.33\%} & -21.5\%\\
\hline
\end{tabularx}
\end{table*}

\section{EXPERIMENTS}
In this section, we conduct experiments to evaluate the effectiveness of AutoML searching for large capacity models for CTR and CVR prediction tasks. Our experiments have the following objectives:
\begin{enumerate}
    \item Assess the end-to-end model accuracy and efficiency improvement of DHEN~\cite{zhang2022dhen} architectures discovered by AutoML compared to manual designs.
    \item Compare the search efficiency between the Predictor-based RL searcher, random searcher and Bayesian optimization to demonstrate the effectiveness of our searching algorithm in web-scale applications.
\end{enumerate}

\subsection{Experiment Setup}

\subsubsection{Search Space Definition}\hfill\vspace{0.5em}
\label{sec:search_space}

In this work, we focus on searching over a Deep Hierarchical Ensemble Network (DHEN) based ranking model. 
Most state-of-the-art high-performance model architectures 
use a deep stacking structure. Here, the deep stacking structures are usually composed of a repeating block containing the same interaction. DHEN also follows this stacking strategy but at the same time builds a novel hierarchical ensemble framework to capture the correlations of multiple interaction modules. It uses the same feature processing layer found in DLRM \cite{naumov2019deep} as feature processing layer. Meanwhile, it proposes a novel hierarchical ensemble framework that includes multiple types of interaction modules and their correlations. Conceptually, a deep hierarchical ensemble network can be described as a deep, fully connected interaction module network, which is analogous to a deep neural network with fully connected neurons. The main goal of hierarchical ensemble is to capture the correlation of the interaction modules. In this work, we largely follow the original DHEN work and consider five types of interaction modules: AdvancedDLRM \cite{mudigere2022software}, self-attention, Linear, Deep Cross Net \cite{wang2017deep}, and Convolution. However, due to the potential overlapping with these modules, we also consider the options of removing them in our search space. In summary, our AutoML search space is composed of the following core components:
\begin{enumerate}
    \item The number of DHEN layers.
    \item Enablement of each interaction module: a binary decision for each interaction module that applies to all layers, e.g., whether to use/not use a particular interaction module for all DHEN layer. 
    \item Model architecture selection: architecture selection for all interaction modules, e.g., kernel size for convolution, or size of weight matrix for linear module,  etc. Again, the architecture selections apply to all DHEN layers. 
    \item Hyperparameter selection: learning rate, weight initialization scales for the weight matrix and layer normalization.
\end{enumerate}

\subsubsection{Searcher hyper-parameter settings}\hfill\vspace{0.5em}

Regarding the hyperparameter settings of the searcher, for the predictor part, we use a four-layer MultiTaskMLP with an architecture [input dim, 50, 50, 2] and 0.5 dropout rate for each layer. The first two layers are a shared architecture, while the last two layers use separate networks for different tasks. Throughout the entire experiment, we use NE gain and FLOPs as the two tasks for metrics, so the output dimension is 2. For the loss function, we use pairwise ranking loss, setting the margin to 0.001. For ensemble modeling, we use 10 individual models to predict together. We use the AdamW algorithm for the optimizer, with a learning rate of 0.001 and a weight decay of 0.005. The learning rate for the RL optimizer is set to 0.01, with an anneal rate of 0.9997, and both the start and min temperature are set at 1.0. Additionally, each time we sample, we use 3 different random seeds to generate top models.
More importantly, all those hyperparameters can be tuned by cross-validation using dataset collected.
As the predictor is lightweight and the RL search is efficient, the cost of hyperparameter tuning is negligible compared with training a ranking model.

\subsubsection{Training hardware}\hfill\vspace{0.5em}

To enable efficient search, we take advantage of the ZionEX fully synchronous training system detailed in \cite{mudigere2022software}. At a high level, the ZionEX system groups 16 hosts into a "supernode", called a pod, which contains 128 A100 GPUs (8 per host) with a total HBM capacity of 5TB and 40PF/s BF16 compute capability. Within a host, each GPU is connected through NVLink, and each host in a pod is then connected with a high bandwidth network of up to 200GB/s, shared with 8 GPUs. With ZionEX, we enable efficient and scalable training of DHEN in our cluster \cite{zhang2022dhen}.
\subsection{Experiment Results}


\begin{figure}
  \centering
  \begin{subfigure}{\linewidth}
    \centering
    \includegraphics[width=\linewidth]{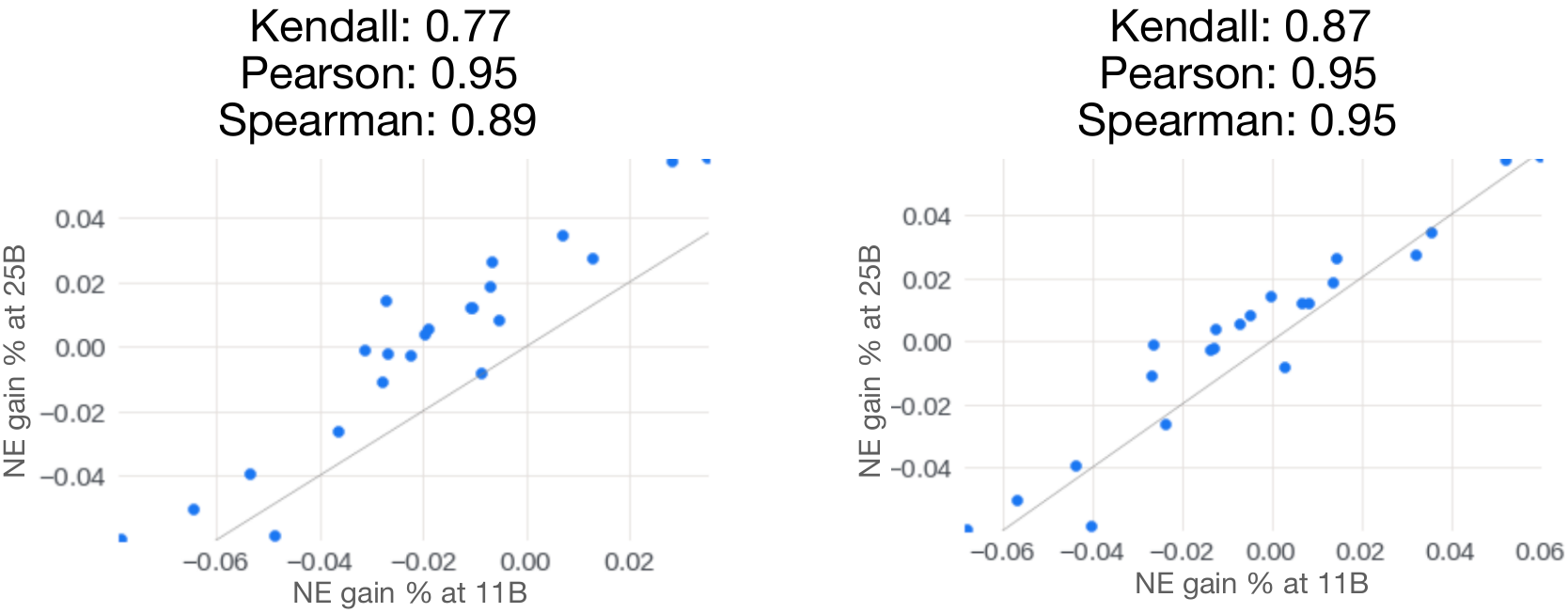}
    \caption{Ranking correlation w/ and w/o curve trends}
  \label{fig:rank_corr_1}
  \end{subfigure}
  
  \begin{subfigure}{\linewidth}
    \centering
    \includegraphics[width=\linewidth]{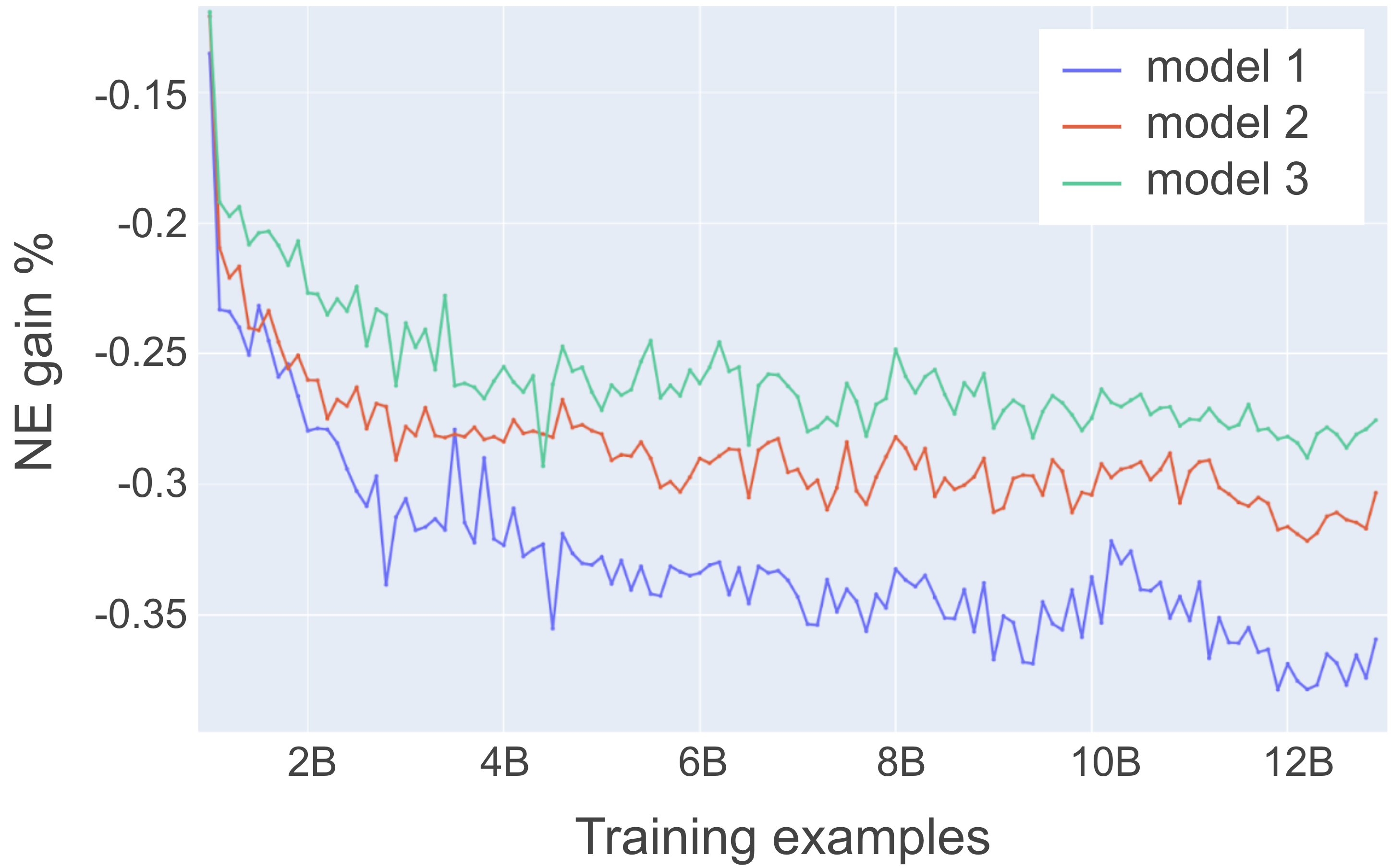}
    \caption{NE learning curves of discovered models using predicted long-term NE}
  \label{fig:rank_corr_2}
  \end{subfigure}

  \caption{Discover model architectures based on both NE and curve trends in CVR scale-up}
  
\end{figure}

\subsubsection{Ranking correlation study}\hfill\vspace{0.5em}

To improve searching efficiency, we applied low-fidelity evaluation in both CTR and CVR scale-up productions to reduce the compute cost (GPU time) required to evaluate each model architectures during the search process. The left-hand-side plot of Figure~\ref{fig:rank_corr_1} shows the results of ranking correlation between low-fidelity evaluation and long-term training evaluation. For example, we typically train a CVR model with 25B (``B'' is for ``billion'') data and expect the NE learning curve to sufficiently converge to a stable value; however, we found through ranking correlation analysis that the NE ranking at 11B data is highly correlated with the results at 25B data (with Kendall Tau $0.77$). Based on this observation, we propose to only use 11B data to evaluate models which can help us reduce compute resources by more than $50\%$ (of the original cost). 

As discussed in Section~\ref{sec:train_pred}, incorporating curve trends to predict long-term NE can further improve ranking correlation. This is demonstrated by the right-hand-side plot in Figure \ref{fig:rank_corr_1} where we see Kendall Tau improve from $0.77$ to $0.87$. Putting it all together, we applied these proposed techniques to discover three top CVR models in Table \ref{tab:search_results} and show their learning curves in Figure \ref{fig:rank_corr_2}.
\begin{figure}
  \centering
  \begin{subfigure}{\linewidth}
    \centering
    \includegraphics[width=\linewidth]{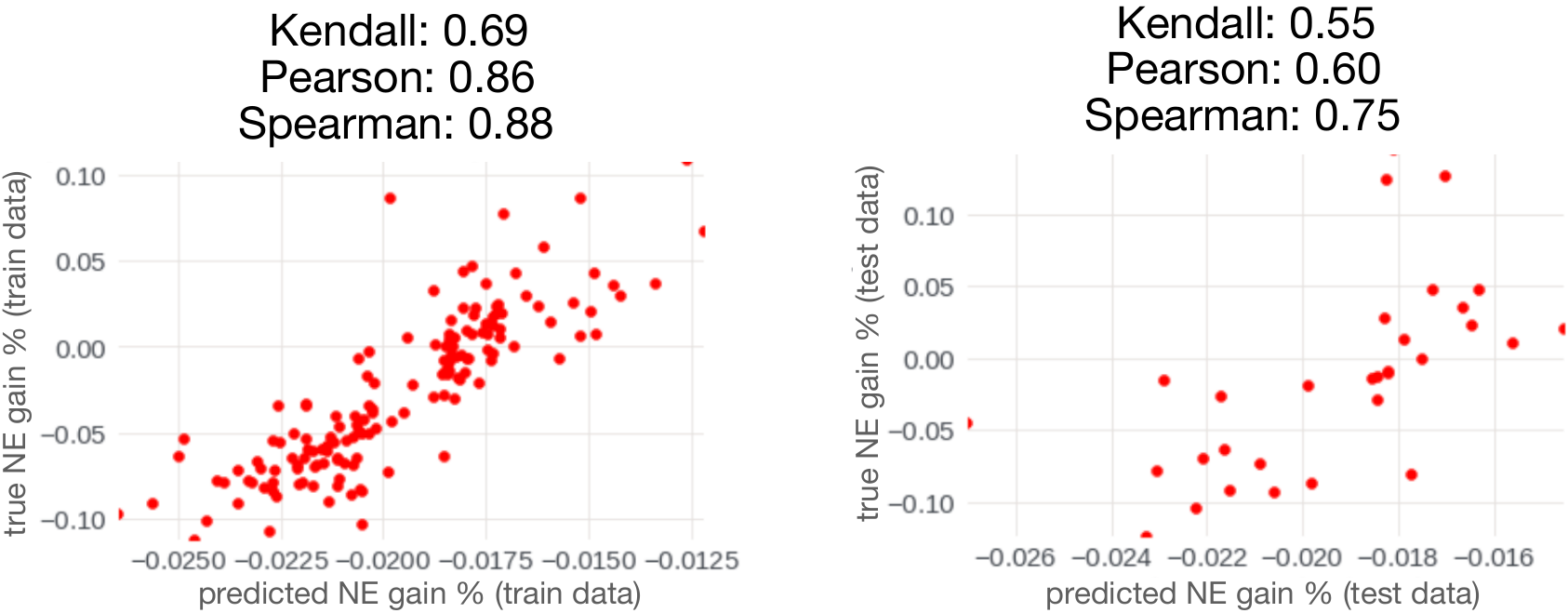}
    \caption{w/ pairwise ranking loss Eq. (\ref{eq:pairwise_rank})}
  \end{subfigure}
  
  \begin{subfigure}{\linewidth}
    \centering
    \includegraphics[width=\linewidth]{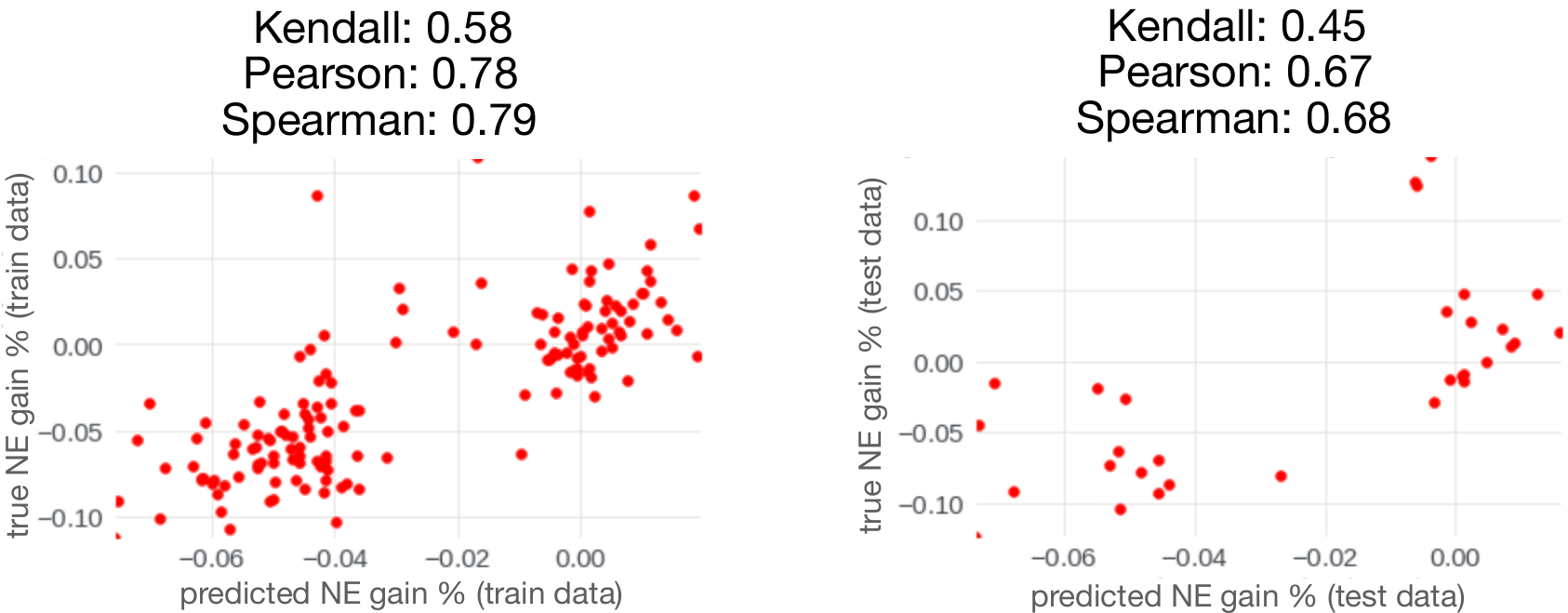}
    \caption{w/ MSE}
  \end{subfigure}

  \caption{Predictor rank correlation between pairwise ranking loss and MSE}
  \label{fig:loss}
\end{figure}

\subsubsection{Ablation study of pairwise ranking loss}\hfill\vspace{0.5em}

We also compared the impact of different types of loss functions on the performance of the predictor. Specifically, we used pairwise ranking loss as Eq. (\ref{eq:pairwise_rank}) and MSE. Through our experiments, we found that although MSE can make the predictor's estimation of NE and FLOPs closer to their true values, when considering the rank correlation of sampled models on real data, pairwise ranking loss obviously performs better than the former (Figure \ref{fig:loss}). This also allows us to more efficiently find the true top models.

\begin{figure}
  \centering
  \begin{subfigure}{\linewidth}
    \centering
    \includegraphics[width=\linewidth]{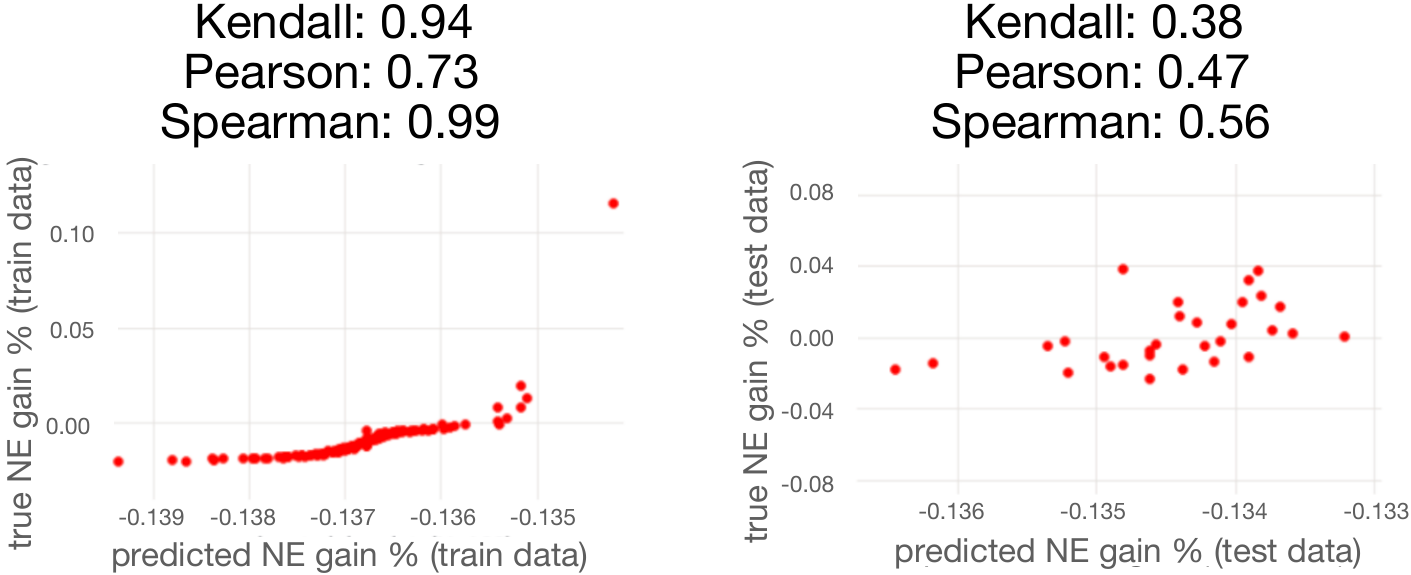}
    \caption{w/o ensemble modeling}
  \end{subfigure}
  
  \begin{subfigure}{\linewidth}
    \centering
    \includegraphics[width=\linewidth]{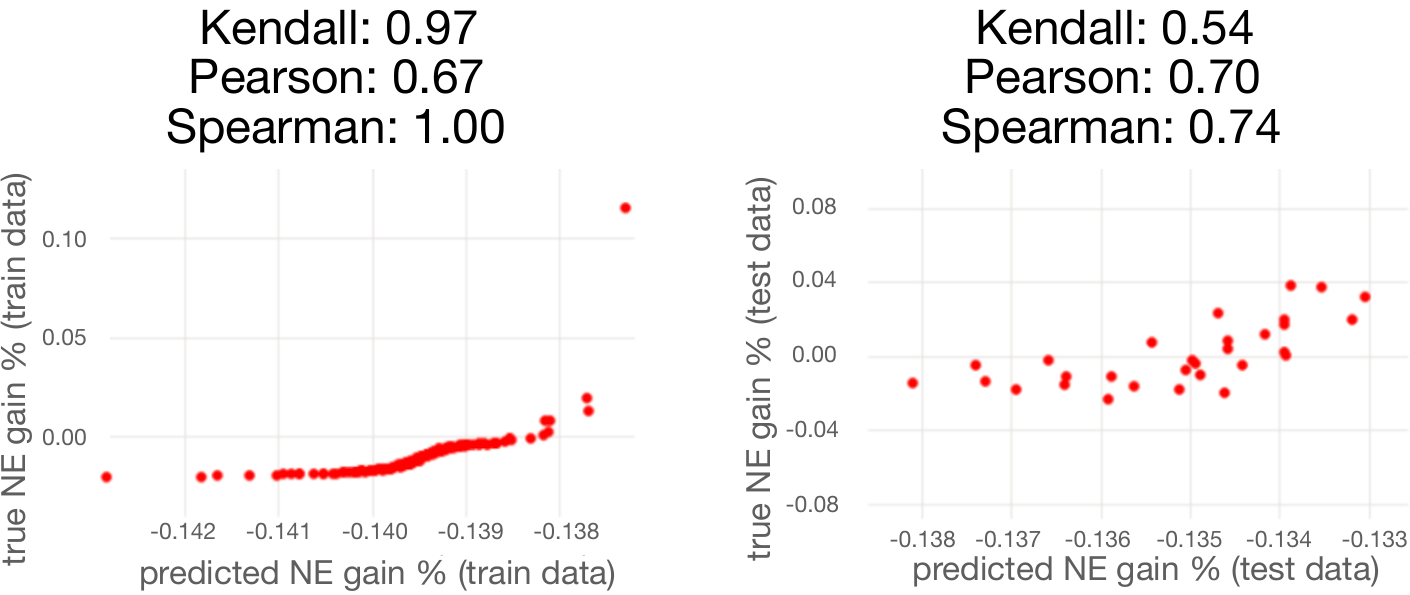}
    \caption{w/ ensemble modeling}
  \end{subfigure}

  \caption{Predictor accuracy performance w/ and w/o ensemble modeling}
  \label{fig:ensemble}
\end{figure}

\subsubsection{Ablation study of ensemble modeling}\hfill\vspace{0.5em}

Due to the tight production schedule and limited computational resources, a predictor realistically only have access to a very limited amount of training data. In real-world cases, we found ensemble modeling can significantly improve a predictor's generalization capacity when training data is very limited. An example is illustrated in Figure \ref{fig:ensemble} where we collected 190 historical model evaluations and divide them into a training set and a validation set with 160:30 split. When using a single model to build the predictor, the ranking correlation between the predictor's predictions on the validation set and the ground truth is much lower than the results on the training set, indicating poor generalization performance. In contrast, when we use 10 models to build the predictor through ensemble modeling and make prediction using Eq.~\eqref{eq:ensemble}, the ranking correlation between the predicted values on the validation set and the ground truth improve significantly. Since the MLP model we used to build predictor are very small model, the additional computational cost introduced by ensemble modeling can be considered negligible.
\subsubsection{Inference FLOPs and QPS correlation study}\hfill\vspace{0.5em}

\begin{figure}
  \centering
  \includegraphics[width=0.8\linewidth]{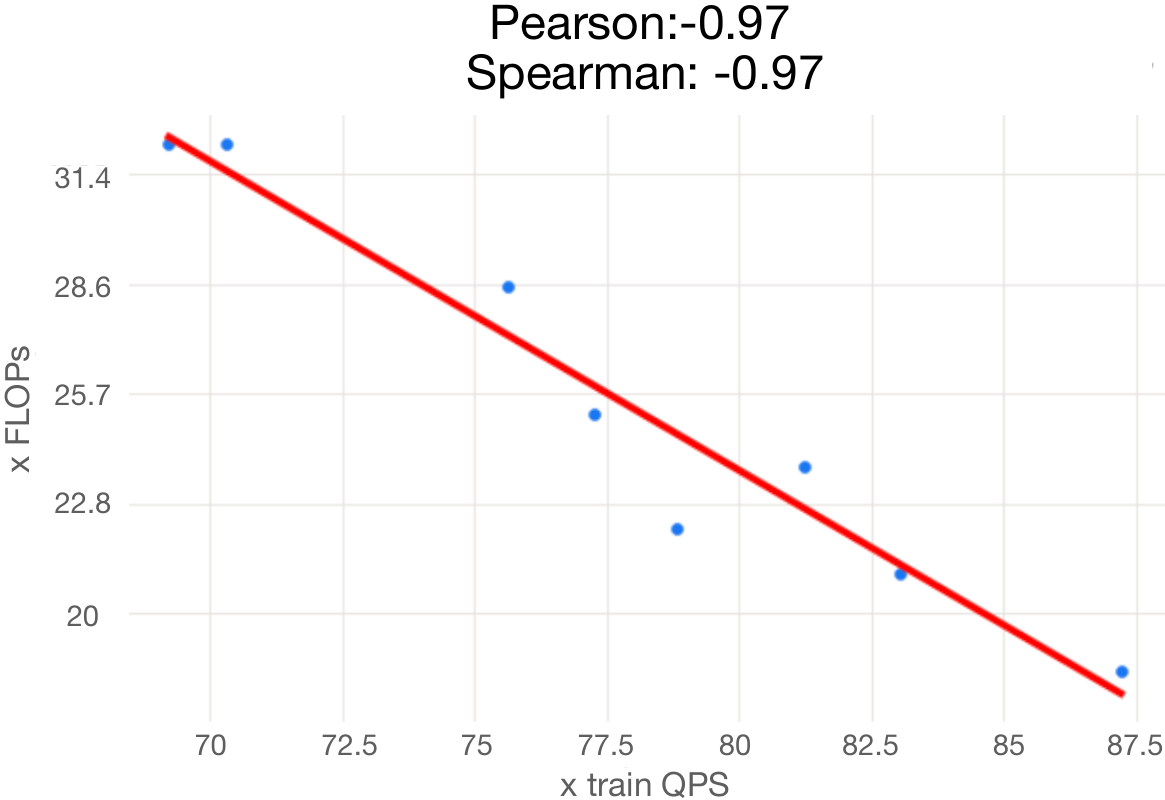}
  \caption{Correlation between inference FLOPs and train QPS}
  \label{fig:flops_qps}
\end{figure}

In terms of hardware infra cost, QPS is the gold metric used at Meta to measure hardware cost instead of FLOPs. However, our training clusters have different GPU types which produce various QPS values for the same model, making the cost incomparable. Moreover, to utilize all the available computing resource, we should not restrict AutoML model evaluation to run on the same specific training hardware type. This makes using QPS as cost metric challenging in real-world AutoML search. To overcome this, we performed study and show that a model's inference FLOPs (in our search space) is highly correlated with train QPS as shown in Figure~\ref{fig:flops_qps} given the same hardware. Based on this observation, we use FLOPs as the cost metric in this work without sacrificing the ability to search models with high QPS.

\begin{figure}
  \centering
  \includegraphics[width=\linewidth]{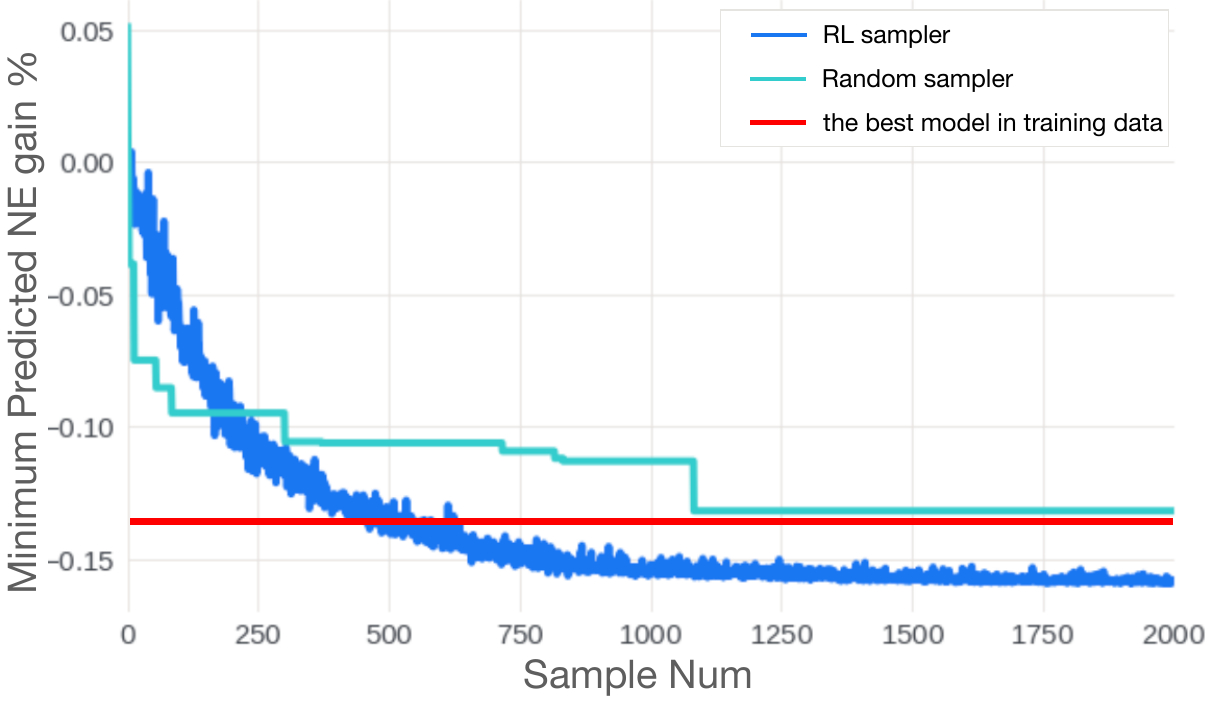}
  \caption{RL versus random sampling efficiency}
  \label{fig:sampler}
\end{figure}

\begin{figure}
  \centering
  \includegraphics[width=\linewidth]{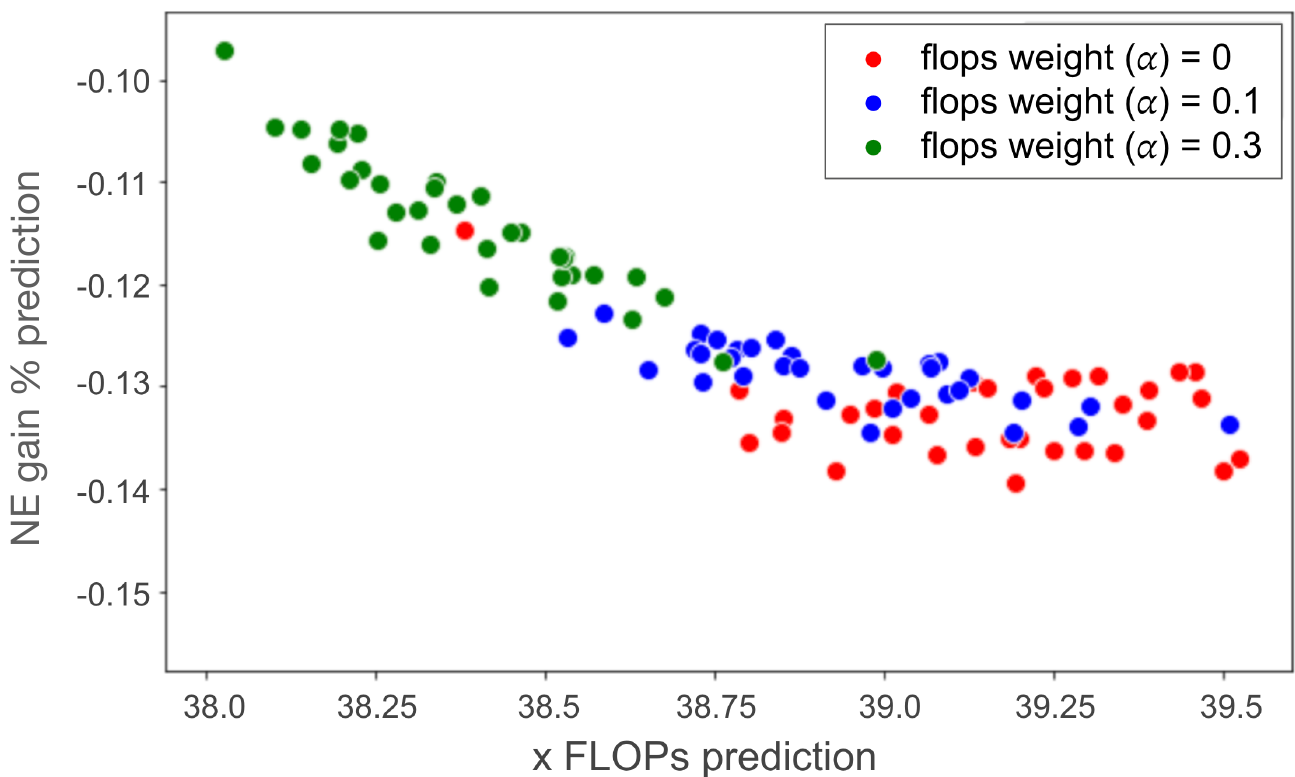}
  \caption{the predicted training NE gain and inference FLOPs of the models identified by RL sampler}
  \label{fig:flops}
\end{figure}

\subsubsection{Ablation study of RL efficient search}\hfill\vspace{0.5em}

We demonstrate the sampling efficiency of RL sampler and compare it to a random sampler. In Figure \ref{fig:sampler},
we can see that the predicted NE of the RL sampled model converge within $2000$ steps and only takes several minutes on a single GPU machine. The final converged NE is lower (better) than both the best model in training data (indicated by the red curve in Figure \ref{fig:sampler}) and the predicted NE of a random sampler under the same sampling steps which further validates the effectiveness of the RL sampler. The highly efficient and effective RL sampler allows us to iterate multiple rounds of RL search, each with a different random seed that leads to the discovery of a different local optimal models. This can help with global exploration as well as finding a diverse set of strong candidates.

Next, to find more promising models with different cost levels, we consider sampling models with cost constraints. Through a study on the correlation among training QPS, inference QPS, and inference FLOPs, we found that these three metrics are highly correlated in CTR and CVR models. Therefore, in AutoML searching, we built a multi-task MLP model to predict training NE and inference FLOPs, and used the weighted sum of the output as the reward in the RL sampler. This helped the predictor effectively search for candidate models at different infra cost levels through adjusting the weighting coefficient $\alpha$. Consequently, in the iteration of the predictor-based RL searcher, we can obtain promising candidates with different FLOPs levels. 
As illustrated in Figure \ref{fig:flops}, we can explore model architectures of various sizes by adjusting the weight associated with the cost constraint in the RL's reward, referred to as the "flops weight". When setting flops weight = 0, it signifies our sole emphasis on NE gain brought about by the model architectures. In this scenario, the RL sampler will recommend many models that prioritize NE gain even if it means sacrificing model efficiency. Conversely, when setting flops weight to a higher value of 0.3, we take into account the impact on model efficiency while striving to enhance NE. Consequently, the recommended models may exhibit relatively lower NE gains, but their model efficiency is significantly improved. Through the utilization of the RL sampler in conjunction with cost constraints for sampling, we have observed that it facilitates the rapid identification of models tailored to different production requirements (as shown in Table \ref{tab:search_results}).
\subsubsection{End to end  search results}\hfill\vspace{0.5em}

To search for a better ROI model for CTR and CVR prediction, we set up a DHEN search space with 18 decisions based on the discussion in Section~\ref{sec:search_space}. Table~\ref{tab:search_results} shows the NE gain and inference QPS improvement of the AutoML discovered DHEN models and a comparison to the manually designed DHEN model and production baseline for CTR and CVR applications.
We validated that the predictor-based RL searcher, whether with moderate cost increase or large capacity modeling, can identify promising architectures at different infrastructure cost levels. For example, in Instagram CTR application, our method can improve NE gain from -0.07\% (based on manual tuning) to -0.13\% while maintaining on-par model efficiency (measured by inference QPS); additionally, we can discover a model that has on-par NE performance but improves inference QPS by 18.96\%.
In the scale-up production for both Instagram CTR and Ad CVR applications, our method can further improve NE gain to -0.09\% without significant model efficiency regression (compared to manual scale-up models). Due to its high ROI, the AutoML generated Instagram CTR scalu-up model with up to -0.36\% NE gain was selected for large-scale online A/B test and show statistically significant gain after 12 days’ reading. All the search results presented in this work are obtained by only sampling on average one hundred models (``\# samples'' in Table~\ref{tab:search_results}), further proving the efficiency of our AutoML algorithms.

\begin{figure}
  \centering
  \includegraphics[width=\linewidth]{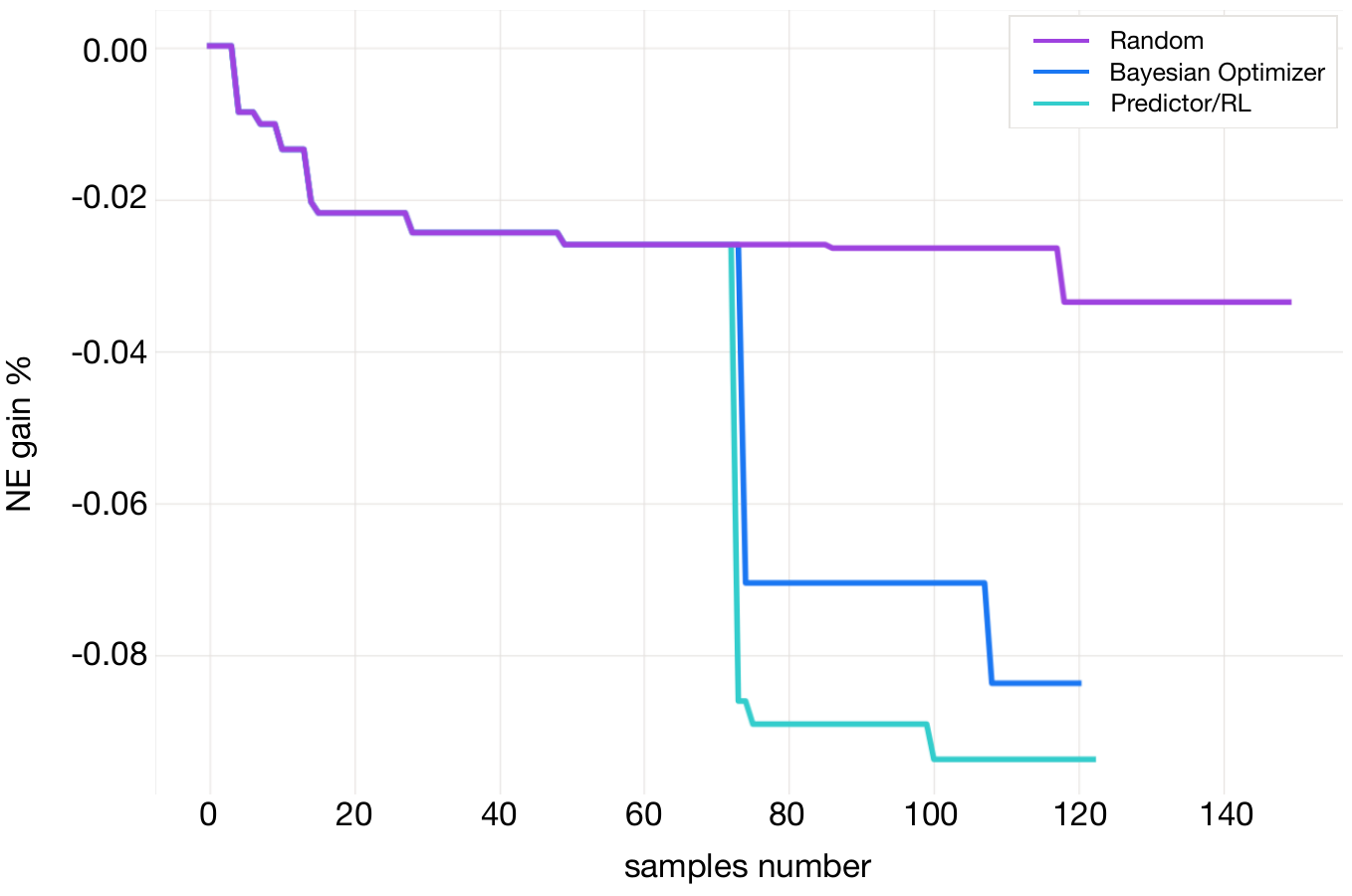}
  \caption{Sample efficiency comparison among different searcher in CVR model}
  \label{fig:vs_rand}
\end{figure}
\begin{table}
\centering
\caption{Different searcher compared to manual design in CVR model}
\label{tab:searcher_comparison}
\begin{tabular}{cc}
\hline
\textbf{Method} & \textbf{window NE gain}\\
\hline
Random & -0.064\%\\
Bayesian Optimizer & -0.082\%\\
Predictor-based RL Searcher & \textbf{-0.093\%}\\
\hline
\end{tabular}
\end{table}

\subsubsection{Predictor-based RL searcher versus Bayesian optimizer versuse Random searcher}\hfill\vspace{0.5em}

Lastly, we conduct a searcher comparison between the proposed predictor-based RL searcher, Bayesian optimizer, and random searcher in a CVR model scale-up problem setting. We compare their search performance over the same compute budget of 150 trials. However, to have a fair comparison we reuse the first 75 trails of random searcher to train predictor-based RL searcher and Bayesian optimizer and then have them generate their remaining 75 candidates. Both predictor-based RL searcher and Bayesian optimizer can be executed in multiple iterations but due to limited compute and time resources in real-world production setting, we typically only perform one iteration of search for both predictor-based RL and BO searchers. We summarize the comparison results in Table \ref{tab:searcher_comparison} where we can see the proposed predictor-based RL searcher greatly improves the model NE performance compared to other searchers. Figure \ref{fig:vs_rand} shows the search performance for each searcher as a function of submitted number of trials.


\section{CONCLUSION}
In this work, we focused on advancing Automated Machine Learning (AutoML) for efficient productionization of Meta-scale ranking system with ever increasing complexity. We introduced an efficient sampling-based AutoML method that is based on a predictor-based searcher. Through innovative techniques such as low-fidelity evaluation, novel metrics for identifying promising models, enhanced generalization with ensemble modeling, and reinforcement learning sampling integration, our method has demonstrated remarkable results. It facilitates large-capacity modeling with superior Return on Investment (ROI) and effectively discovers promising model architectures across different infrastructure cost levels. This work demonstrates the efficiency and adaptability of AutoML in addressing real-world production challenges, ultimately enhancing the impact of machine learning in web-scale applications. 
\bibliographystyle{ACM-Reference-Format}  
\bibliography{main}  

\end{document}